\documentclass[12pt]{article}
\usepackage{amsmath}

\input{tcilatex}

\begin{document}

\title{One Dimensional Asynchronous Cooperative Parrondo's Games}
\author{Zoran Mihailovi\'{c} and Milan Rajkovi\'{c} \\
Vin\v{c}a Institute of Nuclear Sciences, \\
P.O. Box 522, 11001 Belgrade, Serbia}
\maketitle

\begin{abstract}
An analytical result and an algorithm are derived for the probability
distribution of the one-dimensional cooperative Parrondo's games. We show
that winning and the occurrence of the paradox depends on the number of
players. Analytical results are compared to the results of the computer
simulation and to the results based on the mean-field approach.
\end{abstract}

\section{Introduction}

Devised as a pedagogical illustration of the Brownian ratchet mechanism,
Parrondo's games are coin flipping games with an apparently paradoxical
property that alternating plays of two losing games produce a winning game.
In the original setup a player has some capital, which in game $A$ is
increased by one with probability $p$ and decreased by one with probability $%
1-p$ \cite{HATP_1},\cite{HATP_2},\cite{HATP_3}. Game $B$ is more complicated
and the rules are that the probability of winning is $p_{1}$ if the capital
is multiple of $M$ and if it is not, the probability of winning is $p_{2}$.
A bias is introduced into probabilities such that two games have a tendency
to lose (capital is a decreasing function of the number of runs). A third
game, game $A+B$, may be constructed as a random combination of games $A$
and $B$, and this game is in the long run winning (the capital is an
increasing function of the number of runs). New types of Parrondo's game
have been recently introduced in which the feedback is introduced through
spatial neighbor dependence \cite{Toral}, \cite{HA}. These games, termed
cooperative Parrondo's games, rely on the state of player's neighbors
depending whether a player has either lost or won the previous game. Each of 
$N$ players, arranged in a circle, owns a capital $C_{i}(t),i=1,\ldots ,N$,
which evolves by combination of two games. Game $A$ is the same as in the
classical setup, namely the probability of winning and losing is $p^{(A)}$
and $1-p^{(A)}$ respectively. Game $B$ is different from its counterpart in
the original setup since it depends on the state of the neighbors to the
left and to the right of the player. It was shown that alternation of games $%
A$ and $B$, which may be losing or fair when played individually, leads to a
winning outcome. In \cite{Toral}, the evolution of probabilities in games $B$
and $C$ was studied using mean field type equations since it was assumed
that the corresponding exact equations were too complicated to be obtained
analytically. We derive here the exact probability evolution equations
analytically and show excellent agreement with the results based on computer
simulations. We also demonstrate that the existence of the paradox depends
on the number of players, when a set of probabilities introduced in \cite%
{Toral} is used. For arbitrary set of probabilities, we establish the
constraints that prevent the occurrence of the paradox when games are played
by three players. The same approach may be applied to arbitrary number of
players however with considerable increase in complexity of expressions. The
paper is organized as follows: Following a short presentation of essential
rules of the games in section 2, we derive the probability transition matrix
and the corresponding probability evolution equation in section 3.
Stationary probability distribution and constraints in the form of
inequalities for the paradox to occur are derived in section 4 and finally
we compare analytical results with computer simulation results and the mean
field approach of Toral \cite{Toral} in section 5. We conclude with
suggestions about possible new directions and applications of this game.

\section{Features of the Game}

Each player may be in one of two states: state $0$ (``loser'') or state $1$
(``winner''). The state of the whole ensemble of $N$ players may be
represented as a binary string $s=(s_{1},...s_{N}),$ $s_{i}\in (0,1)$ of
length $N$, or equivalently as state \textbf{s }in decimal notation. We also
assume periodic boundary conditions, i.e. $s_{N+1}=s_{1}$. To each state
corresponds a vector, equivalent to a basis vector $\left| s\right\rangle $
in $M=2^{N}$ dimensional state space $S_{M}$%
\begin{equation}
S_{M}=\left\{ \left| s\right\rangle \text{\TEXTsymbol{\vert}}%
s=0,1,...M-1\right\} .  \label{1}
\end{equation}

For example, state $(011)$ is equivalent to state \textbf{3}, and the
corresponding vector is $\left| 3\right\rangle =\left[ 00010000\right] ^{T}$
, while state $(111)$ is equivalent to state \textbf{7} and the
corresponding vector is $\left| 7\right\rangle =\left[ 00000001\right] ^{T}$%
. Game $A$ is the same as in the classical setup, while probabilities of
winning in game B depend on the present state of left and right neighbors,
denoted as a pair $\ (s_{k}-1$ $s_{k+1})$, and with player at position $k$
are given by:

$\cdot$
$p_{0}^{(B)}$ when $(s_{k-1}$ $s_{k+1})$ $=(00)$, in decimal notation 
\textbf{0},

$\cdot$
$p_{1}^{(B)}$ when $(s_{k-1}$ $s_{k+1})=(01)$, in decimal notation \textbf{1}%
,

$\cdot$
$p_{2}^{(B)}$ when $(s_{k-1}$ $s_{k+1})=(10)$, in decimal notation \textbf{2,%
}

$\cdot$
$p_{3}^{(B)}$ when $(s_{k-1}$ $s_{k+1})=(11)$, in decimal notation \textbf{3}%
.

Winning or losing in any particular game leaves a player in state $1$
(``winner'') or $0$ (``loser'') respectively, until he gets a random chance
to play again. Capital $C(t)$ is a function of the ensemble that is
incremented by $1$ or decremented by $1$ if one of the players wins or loses
respectively. Following a play by one of the players, the state of the
ensemble has changed from a state $s(t)$ at time $t$ to a state $s(t+1)$ at
time $t+1$. If the probability that an ensemble in state $s(t)$ (or $\left|
s(t)\right\rangle $\ ) is $\pi _{s}(t)$, then the probability distribution $%
p(t)$ at time $t$ is:%
\begin{equation}
\left| \pi (t)\right\rangle =\sum\nolimits_{s=0}^{M-1}\pi _{s}(t)\text{ }%
\left| s\right\rangle ,  \label{2}
\end{equation}%
while the corresponding probability distribution evolution equation is%
\begin{equation}
\left| \pi (t+1)\right\rangle =\text{ }\mathcal{T}\text{ }\left| \pi
(t)\right\rangle  \label{3}
\end{equation}%
It was reported in \cite{Toral} that the combination of losing games $A$ and 
$B$, leads to a winning result for the set of probabilities $%
(p_{0}=1,p_{1}=p_{2}=0.16,p_{3}=0.7)$. Numerical simulations show that this
set is not unique and in subsequent sections we discuss constraints that
determine this paradoxical result, however in further exposition we will
adhere exclusively to this set.

\section{Probability Transition Matrix}

The analysis is performed via discrete time Markov chains (DTMCs) and we
first derive the probability transition matrix for game $B$. Since at each
moment of time only one player plays and therefore changes state, the change
may be represented by a Hamming distance between the initial ($i$) and the
final ($f$) state of player $k$ is defined as:

\begin{equation}
d_{H}=\dsum\nolimits_{k=1}^{N}\left| i_{k}-f_{k}\right| .  \label{4}
\end{equation}%
Clearly, player may either remain in the same state as before the play ($%
d_{k}^{H}=|i_{k}-f_{k}|=0$) or change his state ($d_{k}^{H}=|i_{k}-f_{k}|=1$%
). We will consider each case separately.

\textbf{Case 1: }$d_{H}=0$

If the ensemble is in state \textbf{s}, then the k-th player is in state $%
s_{k}$. State \textbf{s} of the ensemble also defines the neighborhood of
the $k$-th player, a pair $\eta _{k}=(s_{k-1},$ $s_{k+1})$which in turn
determines the probability of winning. Since the ensemble initially in state 
\textbf{i} may switch to state \textbf{f} = \textbf{i} in one of $N$
different ways as a result of one of the players switching from state $i_{k}$
to state $f_{k}=i_{k}$, the probability of transition is in this case equal
to the sum of probabilities of independent events

\begin{equation}
T_{fi}=w(i\rightarrow f)=\frac{1}{N}\dsum\nolimits_{k=1}^{N}w(i_{k},f_{k}),
\label{5}
\end{equation}%
where probabilities $w$ depend on whether state $f_{k}$ is $1$(winning) or $%
0 $ (losing), and upon the probability of winning, i.e.

\begin{equation}
w(i_{k},f_{k})=\left\{ 
\begin{array}{cc}
1-p_{\eta _{k}}^{(B)} & f_{k}=0 \\ 
p_{\eta _{k}}^{(B)} & f_{k}=1%
\end{array}%
\right.  \label{6}
\end{equation}%
\textbf{Case 1: }$d_{H}=1$

In this case $k$-th player switches from state $i_{k}$ to state $f_{k}$ , ($%
i_{k}$ $\neq $ $f_{k}$ ), with probability

\begin{equation}
T_{fi}=\frac{1}{N}w(i_{k},f_{k}).  \label{7}
\end{equation}

\subsection{Probability transition matrix for $N=3$}

As an example we consider an ensemble of three players $(N=3)$ playing game $%
B$, and let us assume that the ensemble is in state \textbf{s} = $(011)$= 
\textbf{3}. Possible transitions and the corresponding probabilities are:

\begin{equation}
\begin{array}{l}
T_{33}=w(011\rightarrow 011)=w(\mathbf{3}\rightarrow \mathbf{3})=\frac{1}{3}%
\left[ (1-p_{3})+p_{1}+p_{2}\right]  \\ 
T_{31}=w(011\rightarrow 001)=w(\mathbf{3}\rightarrow \mathbf{1})=\frac{1}{3}%
(1-p_{1}) \\ 
T_{23}=w(011\rightarrow 010)=w(\mathbf{3}\rightarrow \mathbf{2})=\frac{1}{3}%
(1-p_{2}) \\ 
T_{73}=w(011\rightarrow 111)=w(\mathbf{3}\rightarrow \mathbf{7})=\frac{1}{3}%
p_{3}%
\end{array}
\label{8}
\end{equation}%
Explicitly, probability $T_{13}$ represents the probability that second
player, in state $s_{2}=1$, switches to state $f_{2}=0$. This probability is
equal to the product of the probability that it is this player's turn to
play i.e. $1/3$, and the probability that this player loses, i.e. switches
to state $f_{2}=0$. The neighborhood of the second player, $(s_{1\text{ }%
}s_{3})=(0$ $1)=1$, determines the choice $p_{1}$. Similarly, transition $%
T_{33}$ may take place when either one of the three players switches to a
same state, hence the corresponding probability is the sum shown in the
first expression of \ref{8}. Finally, using%
\begin{eqnarray*}
p_{i}^{1} &=&1-p_{i},\text{ }i=0,1,2,3 \\
Q_{1} &=&{1-p}_{_{3}}{+p}_{_{1}}{+p}_{_{2}}\text{ \ and} \\
Q_{2} &=&{2-p}_{_{2}}{-p}_{_{1}}{+p}_{_{0}}
\end{eqnarray*}%
the transition matrix for game $B$ $(N=3)$ is:%
\begin{equation}
\mathcal{T}^{(B)}=\left( 
\begin{array}{cccccccc}
{3p}_{_{0}}^{1} & {p}_{_{0}}^{1} & {p}_{_{0}}^{1} & {0} & {p}_{_{0}}^{1} & {0%
} & {0} & {0} \\ 
{p}_{0} & Q_{{2}} & {0} & {p}_{_{1}}^{1} & {0} & {p}_{_{2}}^{1} & {0} & {0}
\\ 
{p}_{0} & {0} & Q_{{2}} & {p}_{_{2}}^{1} & {0} & {0} & {p}_{_{1}}^{1} & {0}
\\ 
{0} & {p}_{_{1}} & {p}_{_{2}} & Q_{1} & {0} & {0} & {0} & {p}_{_{3}}^{1} \\ 
{p}_{0} & {0} & {0} & {0} & Q_{{2}} & {p}_{_{1}}^{1} & {p}_{_{2}}^{1} & {0}
\\ 
{0} & {p}_{_{2}} & {0} & {0} & {p}_{_{1}} & Q_{1} & {0} & {p}_{_{3}}^{1} \\ 
{0} & {0} & {p}_{_{1}} & {0} & {p}_{_{2}} & {0} & Q_{1} & {p}_{_{3}}^{1} \\ 
{0} & {0} & {0} & {p}_{_{3}} & {0} & {p}_{_{3}} & {p}_{_{3}} & {3p}_{_{3}}%
\end{array}%
\right)   \label{9}
\end{equation}%
Corresponding matrix for game $A$ may be easily obtained by performing the
following replacement: $p_{\eta }^{(B)}\rightarrow p^{(A)}$ for each $p\in $ 
$\left\{ 0,1,2,3\right\} $.

Furthermore, we introduce a vector of the capital $\left| C\right\rangle $%
whose components represent the capital generated by each ensemble state.
Elements $C_{s}$ of this vector represent normalized capital generated by
that specific state (equal to the sum of all winning and losing individual
states in a given ensemble state). Naturally, player state $0$ generates
capital $-1$, while state $1$ generates capital $+1$. Explicitly,%
\begin{equation}
C_{s}=\frac{1}{N}\dsum\nolimits_{i=1}^{N}(-1)^{s_{i}+1},\text{ so that }%
C_{s}\in \left[ -1,1\right] .  \label{10}
\end{equation}%
In other words, elements of $\left| C\right\rangle $ are average values of
the capital generated by each ensemble state separately. For example for $%
N=3 $, the vector of the capital is:%
\begin{equation}
\left| C\right\rangle =(1/3)\left[ -3\text{ }-1\text{ }-1\text{ }1\text{ }-1%
\text{ }1\text{ }1\text{ }3\right] ^{T}.  \label{11}
\end{equation}%
In the above expression the third vector element corresponding to the state $%
(010)=$\textbf{2} is equal to $C_{2}=(1/3)((-1)+(+1)+(-1))=(1/3)(-1)$. This
also implies that the ensemble switching from a state $s(t)$ to a state $%
s(t+1)=(010)=$ $\left| 2\right\rangle $, under the assumption that such a
transition is possible, generates average capital $\left\langle
C(t+1)\right\rangle $=$\left\langle C\text{ }|\text{ }2\right\rangle =-1/3$.
Furthermore, an ensemble remaining in state $\left| 2\right\rangle $
throughout its temporal evolution would in the aver-age generate capital $%
(-1/3)$ in each turn of the game. Hence the capital generated by the
ensemble is

\begin{equation}
\left\langle C\right\rangle =\left\langle C\text{ \TEXTsymbol{\vert} }\pi
\right\rangle ,  \label{12}
\end{equation}%
where $\left\langle C\right\rangle $\ denotes the average value of the
generated capital. In order to evaluate the probability for one of the
games, either $B$ or $A+B$ to be winning, it should be noted that:

\begin{eqnarray}
P_{win}+P_{lose} &=&1  \label{13} \\
P_{win}-P_{lose} &=&\left\langle C\right\rangle  \notag
\end{eqnarray}%
where $P_{win}$ and $P_{lose}$\ are probabilities of winning and losing in a
certain game, so that 
\begin{equation}
P_{win}=(1/2)(1+\left\langle C\right\rangle ).  \label{14}
\end{equation}%
This expression implies that condition $P_{win}^{(B)}<1/2$ , i.e. that game $%
B$ is losing, is equivalent to the condition $\left\langle
C^{(B)}\right\rangle $ $<0$, and that $P_{win}^{(A+B)}>1/2,$ i.e. that game $%
A+B$ is winning, is equivalent to $\left\langle C^{(A+B)}\right\rangle $ $>0$%
.

\section{Analysis of the Games}

\subsection{Equilibrium distribution}

The equilibrium (stationary) state occurs when the probability distribution
remains invariant under the action of $\mathcal{T}$, that is, $\left| \pi
(t+1)\right\rangle =$ $\mathcal{T}$ $\left| \pi (t)\right\rangle $ $=$ $%
\left| \pi \right\rangle $. In order to evaluate the probability
distribution in this case, we need to solve $(\mathbf{1}-\mathcal{T})\pi =0$%
. For game $A$, for which there is a probability $p$ for a player to win
(alternatively $(1-p)$ to lose), the stationary distribution is easily
obtained by setting $p_{0}=p_{1}=p_{2}=p_{3}=p$ and reads 
\begin{eqnarray}
\pi ^{(A)} &=&[(1-p)^{3},(1-p)^{2}p,(1-p)^{2}p,(1-p)p^{2},  \label{15} \\
&&(1-p)^{2}p,(1-p)p^{2},(1-p)p^{2},p^{3}]^{T}
\end{eqnarray}%
The probabilities in the above expression may also be readily obtained by
associating to each ensemble state, from \textbf{0} to \textbf{7} in decimal
notation, corresponding probabilities for each player. For game $B$ the
stationary distribution is%
\begin{eqnarray}
\pi ^{(B)} &=&[\frac{(1-p_{0})[2-(p_{1}+p_{2})](1-p_{3})\alpha }{%
p_{0}p_{3}(p_{1}+p_{2})},\text{ }\frac{[2-(p_{1}+p_{2})](1-p_{3})\alpha }{%
p_{0}p_{3}(p_{1}+p_{2})},  \label{16} \\
&&\frac{[2-(p_{1}+p_{2})](1-p_{3})\alpha }{p_{0}p_{3}(p_{1}+p_{2})},\frac{%
\alpha (1-p_{3})}{p_{3}},\frac{[2-(p_{1}+p_{2})](1-p_{3})\alpha }{%
p_{0}p_{3}(p_{1}+p_{2})}  \notag \\
&&\frac{\alpha (1-p_{3})}{p_{3}},\frac{\alpha (1-p_{3})}{p_{3}},\alpha ]^{T},
\notag
\end{eqnarray}%
where

\begin{equation}
\alpha =\frac{1}{\frac{[(1-p_{0})+3p_{3}](1-p_{3})[2-(p_{1}+p_{2})](1-p_{3})%
}{p_{0}p_{3}(p_{1}+p_{2})}-\frac{3(1-p_{3})}{p_{3}}+1}.  \label{17}
\end{equation}%
Substituting the set of probabilities given in \cite{Toral}, $%
p_{0}=1,p_{1}=p_{2}=0.16,p_{3}=0.7$, in the expression for the stationary
distribution for game $B$, we get 
\begin{equation}
\pi ^{(B)}=[0,.24901,.24901,.04743,.24901,.04743,.04743,.11067]^{T}
\label{18}
\end{equation}%
For the randomized game $A+B$, probabilities $p_{i}(i=0,1,2,3)$ for game $B$%
, are replaced with the corresponding probabilities $q_{i}(i=0,1,2,3)$given
by:%
\begin{equation}
q_{i}=\gamma p+(1-\gamma )p_{i},\text{ \ \ \ }(i=0,1,2,3).  \label{19}
\end{equation}%
and where parameter $\gamma $ represents the relative probability of playing
game $A$, where we have assumed the value of one half. The corresponding
stationary distribution for the randomized game, with $p=1/2$, is%
\begin{equation}
\pi ^{(A+B)}=[.06006,.18019,.18019,.08875,.18019,.08875,.08875,.13312]^{T}.
\label{20}
\end{equation}%
Expressions \ref{18} and \ref{20}, although not leading to the paradoxical
result illustrate the probabilities associated with each state in games $B$
and $A+B$ respectively.

\subsection{Constraints of the games}

The probability of winning using the stationary distribution is given by

\begin{equation}
p_{winning}=\left\langle \pi \text{ }|\text{ }\rho \right\rangle =\text{ }%
\dsum\nolimits_{s=0}^{M-1}\pi _{s}\rho _{s},  \label{21}
\end{equation}%
where $\rho _{s}$ is the winning probability in state $\pi _{s}$. Components 
$\rho _{s}$ of vector $\left| \rho \right\rangle $ may be defined using \ref%
{14} or alternatively as,

\begin{equation}
\rho _{s}=\dsum\nolimits_{k=1}^{N}\frac{s_{k}}{N},  \label{22}
\end{equation}%
so that the probability that the ensemble generates a winning outcome when
switching to a state $s=(s_{1},\ldots ,s_{N}),$ is determined by the
fraction of winning players in the ensemble. The summation in the above
expression should actually be over the indices corresponding to the winning
players, however use of all indices is justified since states $0$ do not
contribute to the summation. Since for the paradox to occur we must have $%
P_{win}^{(B)}<1/2$ and $P_{win}^{(A+B)}>1/2$ simultaneously, a simple
calculation yields the following inequality 
\begin{equation}
(p_{1}+p_{2})[1+\frac{1}{2}(p_{0}-p_{3})]>2[1-\frac{1}{4}(p_{0}+3p_{3})].
\label{23}
\end{equation}%
We may also assume that probabilities $p_{1}$ and $p_{2}$ should be equal,
so that finally we get%
\begin{equation}
1>p_{1}>\frac{1-\frac{1}{4}(p_{0}+3p_{3})}{1+\frac{1}{2}(p_{0}-p_{3})}.
\label{24}
\end{equation}
If $p_{0}=1$, and $p_{3}=0.7$ are inserted in the above expression, the
inequality is not satisfied and there is no paradox, justifying the
conclusion based on numerical simulations that there is no paradox for games
played by three players when probabilities are those given by Toral in \cite%
{Toral}. However, for the same number of players and a set of probabilities,
e.g. $p_{0}=0.09,p_{1}=p_{2}=0.52$ and $p_{3}=0.89,$ the paradox exists.

\section{Results}

All results in this section are based on probability values given in \cite%
{Toral}, namely $(p_{0}=1,p_{1}=p_{2}=0.16,p_{3}=0.7,P^{(A)}=0.499).$ First,
in Fig. 1 we present analytical results that show how paradox depends on the
number of players when $N\leq 12$. Namely, in this range of $N$ values
paradox exists if $N$ is different from $3,4,7$ or $8$. Numerical
simulations, shown in Fig. 2, indicate that the paradox occurs if the number
of players is greater than $8$ and no exceptions are noticed up to $1000$
players. The mean-field equation for the evolution of the common probability
in game $B$ 
\begin{equation}
P^{(B)}(t+1)=(1-\text{ }P^{(B)}(t))^{2}p_{0}+P^{(B)}(t)(1-\text{ }%
P^{(B)}(t))(p_{1}+p_{2})+P^{(B)}(t)^{2}p_{3},  \label{25}
\end{equation}%
in which each term reflects one of the four possibilities given in the
definition of the game, introduced in \cite{Toral}, naturally yields results
irrespective of the number of players.

As a comparative illustration of results based on all three approaches, we
show in Fig. 3 the average capital per turn as a function of the probability 
$p_{3}$ of game $B$ played by five players. Analytical and numerical
simulation results show perfect agreement while results based of the mean
field equation show very good agreement only in the (approximate) range $%
0.16<p_{3}<0.83.$ For values of $p_{3}$ below $0.16$ the capital based on
mean field calculations shows that there is no stable solution of the
iterative mean field equation. Namely, mean field equation in this area
switches to one of the two solutions at each successive step of calculations
while it deviates considerably from the analytical and numerical results for 
$p_{3}>0.83$. We emphasize that due to the very nature of the mean-field
approach this large discrepancy for low and high values of $p_{3}$ remains
for arbitrary number of players. However the dependence of the average
capital as function of other probabilities, $p_{0},p_{1}$ and $p_{2}$ is
considerably better, as presented in Fig. 4. for the case of $p_{0}$ .

\section{Conclusion}

An algorithm for obtaining analytical expressions for the evolution of the
probabili-ties and constraints of the games is illustrated in the case of
three players. These results may prove to be of important use in
applications to social and economic models where exact results may be
indispensable. Usefulness of these games to social and economic models, and
possibly in biological applications, may be expanded by analyzing games
played by all players simultaneously that we introduce in the follow up
paper \cite{MR}. Also, the approach presented here may be useful in analysis
of games based on spin models. Moreover, we show that the mean-field
approach, although useful when large number of players is involved, may be
quite inaccurate in certain range of probabilities that define spatial
neighbor dependence.

\bigskip

One of the authors (M.R.) wishes to thank F. Marchesoni and P. Hanggi, the
organizers of the workshop on ''Stochastic Systems: from Randomness to
Complexity'' (Erice, July 26-August 01, 2002), for the opportunity to
receive a first hand information about Parronodo's games from J.M.R.
Parrondo and R. Toral.

This work is financed by Ministry of Science and Technology of Serbia, under
the grant OI 1986.

\bigskip

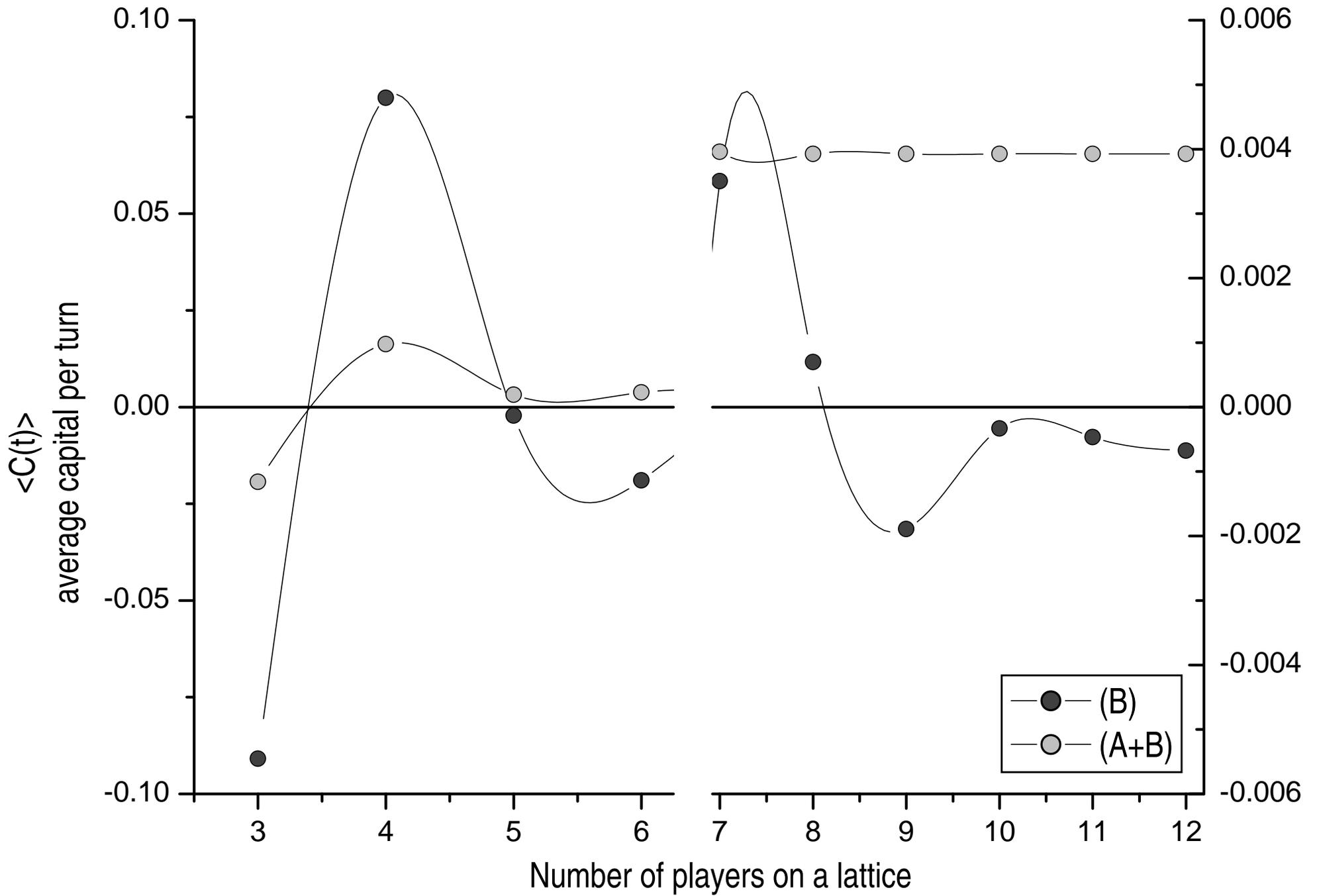
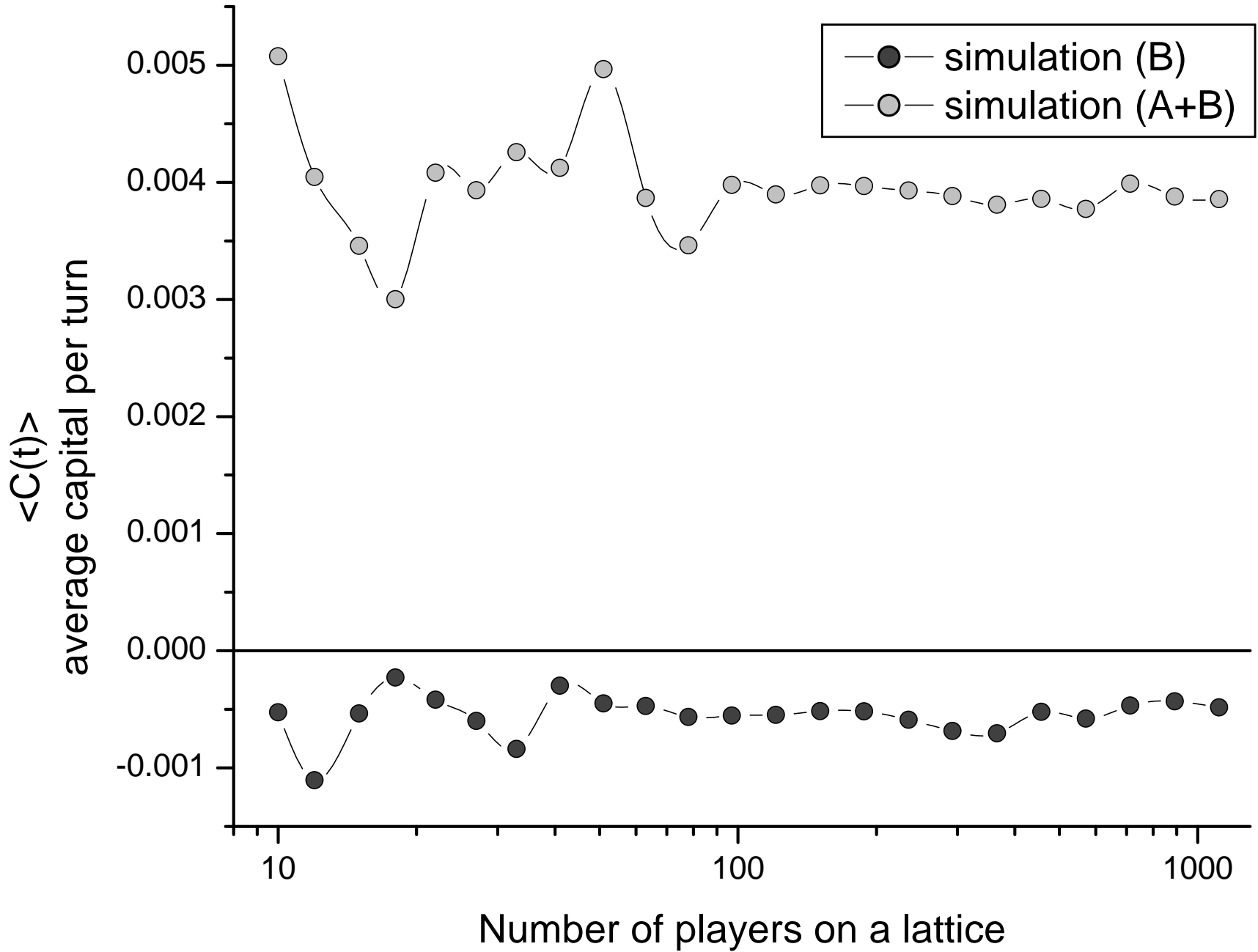
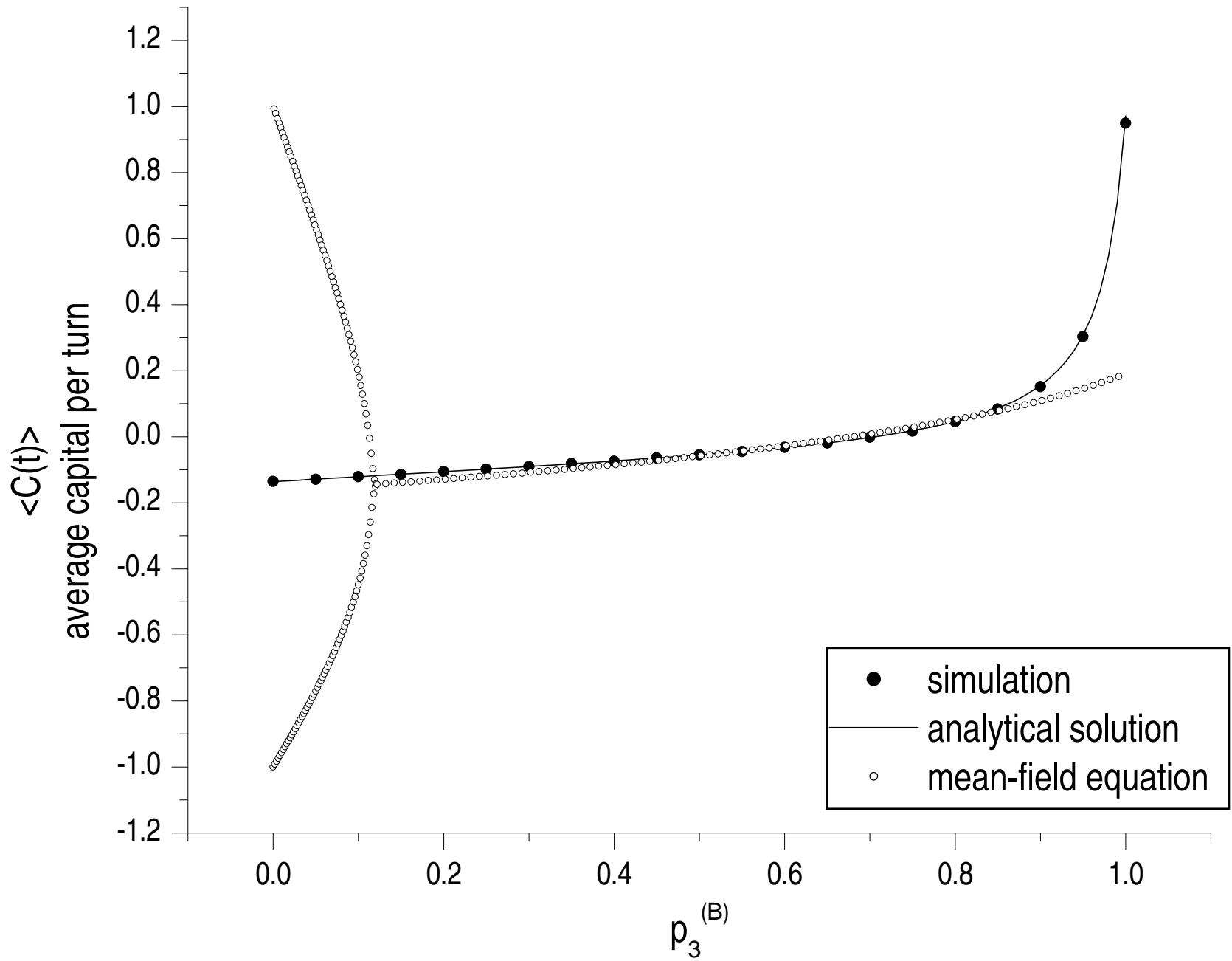
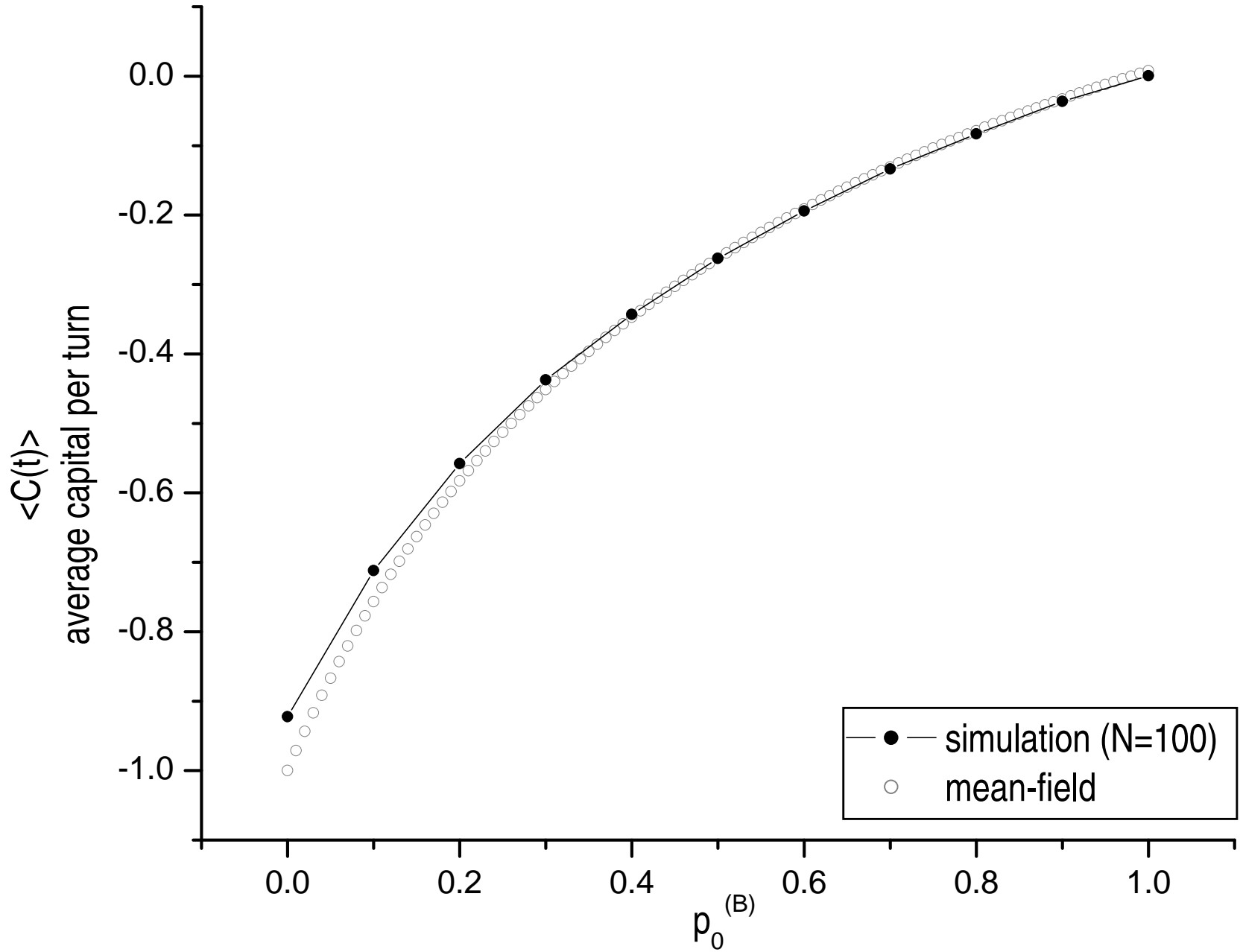
\end{document}


Fig. 1. Averaged capital per turn as a function of the number of players
obtained using analytical expressions for the evolution of the probability
distribution.. For $1\leq N\leq 6$ it is easy to distinguish winning and
losing games, however for $N\geq 7$game $B$ is barely losing and in order to
accentuate that conclusion we use magnified scale for the capital. The
spline fit is performed solely for visualization purposes.

Fig.2. Numerically obtained averaged capital per turn (realization of the
games) as a function of the number of players $(N)$. Capital is averaged
over $1000$ runs of the game, and over $100N$ turns in each run after the
system entered the stationary state.

Fig. 3. Mean-field approach results compared to the analytical and numerical
results for the average capital per turn as a function of $p_{3}^{(B)}$ for $%
N=5$, $(p_{0}=1,p_{1}=p_{2}=0.16)$. Analytical and numerical results show
excellent agreement while mean-field results show considerable divergence
for small and large values of $p_{3}^{(B)}$.

Fig. 4 Average capital per turn of the game as a function of $p_{0}^{(B)}$.
Due to the large number of players only results based on the computer
simulations are compared to the mean-field results. The other probabilities $%
p_{1}=p_{2}=0.16,$ $p_{3}=0.7$ are kept fixed. The graph shows very good
agreement.